\documentclass[a4paper]{article}
\usepackage{amscd,amsmath,amssymb}
\usepackage[usenames, dvipsnames]{color}

\newcommand{\dimg}{{\mbox dim }\mathfrak{g}}

\newcommand{\C}{\mathbb C}

\newcommand{\Z}{\mathbb Z}

\newcommand{\p}[1]{(\ref{#1})}

\newcommand{\halpha}{{\hat \alpha}}
\newcommand{\hbeta}{{\hat \beta}}
\newcommand{\hgamma}{{\hat \gamma}}

\newcommand{\cE}{{ {\cal E}   }}

\newcommand{\Tr}{\textrm{Tr}}
\newcommand{\ad}{{\mathfrak{ad}}}

\newcommand{\be}{\begin{equation}}
\newcommand{\ee}{\end{equation}}
\newcommand{\bea}{\begin{eqnarray}}
\newcommand{\eea}{\end{eqnarray}}

\newcommand{\ba}{\begin{array}} \newcommand{\ea}{\end{array}}

\newcommand{\nn}{\nonumber}
\def\theequation{\arabic{section}.\arabic{equation}}

\begin{document}
\begin{flushright}
\end{flushright}\vspace{1cm}
\begin{center}
	{\Large\bf  The split 5-Casimir operator and the structure of  $\wedge \ad^{\otimes 5}$ }
\end{center}
\vspace{1cm}

\begin{center}
{\Large\bf Alexey P. Isaev${}^{a,b}$ and Sergey O. Krivonos$^{a,c}$}
\end{center}

\vspace{0.2cm}

\begin{center}
	
	\vspace{0.3cm}
	
	{\it
		${}^{a)}$Bogoliubov  Laboratory of Theoretical Physics, JINR,
		141980 Dubna, Russia \\ \vspace{0.2cm}
		${}^{b)}$Faculty of Physics, Lomonosov Moscow State University, 119991 Moscow, Russia\\ \vspace{0.2cm}
		${}^{c)}$Laboratory of Applied Mathematics and Theoretical Physics, TUSUR, Lenin. Ave 40, 634050 Tomsk, Russia}
	
	\vspace{0.5cm}
	
\end{center}
\vspace{2cm}

\begin{abstract}
	\noindent In the present paper, using the  split Casimir operators we have
	found the decomposition of the antisymmetric part of $\ad^{\otimes 5}$. This decomposition contains the representations that appeared in the decomposition of 
	$\ad^{\otimes 4}$ and only one new representation $X_5$. The dimension of this representation has been proposed in \cite{Mac}.  Our decomposition is valid for all Lie algebras.
\end{abstract}

\newpage

\pagenumbering{arabic}
\setcounter{page}{1}
\section{Introduction}
The study of $\ad^{\otimes n}$ up to $n=4$ \cite{Cohen, Landsberg,  AIKM} confirmed the Deligne 
conjecture \cite{Deligne}  concerning Vogel's uniformity \cite{Vogel}. However, the calculations, being more or less reasonable for the exceptional algebras, quickly become very complicated for  semi-simple  Lie algebras of the classical series. Moreover, it seems that no systematic algebraic approach was formulated or, at least, reported. 

In a series of papers \cite{IsPr, AIKM, IsKri, IsKri2}, the use of the split Casimir operator (see e.g. \cite{Book2}) was advocated as the main tool for analyzing the decomposition of powers of  adjoint representations. The (second) split Casimir operator can be defined for any simple complex Lie algebra $\mathfrak{g}$ with the basis elements $X_a$ as
\be\label{splitC}
\C_{(2)} = g^{ab}\, X_a \otimes X_b,
\ee
where $g^{ab}$ is the inverse of the Cartan-Killing metric $g_{ab}$ defined in the standard way as
\be\label{g}
g_{ab} \equiv  \Tr (\ad(X_a)\cdot \ad(X_b)).
\ee
In the simplest case of the second split Casimir operator \p{splitC} it can be further decomposed
into symmetric and anti-symmetric parts:
\be
\C_{(2)} = \left( Sym_{2} \C_{(2)} \right) +  \left( ASym_{2} \C_{(2)} \right) ,
\ee
where the projectors $Sym_{2}$ and $ASym_{2}$ are defined  as
\be\label{Proj2}
Sym_2  =  \frac{1}{2}\left(1+ \sigma_{12}\right) \cE_2,\quad  ASym_2  =  \frac{1}{2}\left(1- \sigma_{12}\right) \cE_2.
\ee
Here,
\be
\sigma_{mn}  =   \delta^{i_m}_{j_n} \delta^{i_n}_{j_m}, \quad  \cE_m =\prod_{k=1}^m \delta^{i_k}_{j_k} .
\ee
In addition, the Split 2-Casimir operator obeys the following identities:
\be
C^{ij}_{il}=C^{il}_{jl} =0, \quad C^{ij}_{ji} =\dimg, \quad C^{ii}_{jj} = -\dimg,
\ee
where $\dimg$ is the dimension of the algebra in question.

One of the consequences of uniformity in application to the split Casimir operators is the existence of a universal form of  traces of their powers as well as the powers of their irreducible parts. For the antisymmetric part of the  Split 2-Casimir operator we have
\bea
Tr(ASym_2) & = &\frac{1}{2} \dimg (\dimg-1) , \nn \\
Tr(ASym_2 \, \C_{(2)}^k) & = & \left(-\frac{1}{2}\right)^k \dimg , \quad \forall k \in \Z_{>0} \label{tr2} .
\eea
The antisymmetric part of $\ad^{\otimes 2}$ includes the following representations \cite{Vogel}
\be\label{dec2}
\wedge \ad^{\otimes 2} = X_1\oplus X_2,
\ee
where
\be
X_1 = \ad \quad\mbox{and} \quad dimX_2  = \frac{1}{2} \dimg \left( \dimg-3\right).
\ee
In \cite{IsPr}, it was shown that, as the consequence of this decomposition, the antisymmetric part of the  split 2-Casimir operator obeys to the following characteristic identity\footnote{Note, that eigenvalues of the split 2-Casimir operator $ \C_{(2)}$ on the representations $X_1$ and $X_2$ are
	equal to $ -\frac{1}{2}$ and $0$, respectively.}:
\be
ASym_2 \, \C_{(2)} \left( \C_{(2)}+\frac{1}{2} \right) =0.
\ee
In the previous papers \cite{IsPr, IsKri, IsKri2}, we used these identities to get decomposition formulas for $\ad^{\otimes n}, \; n= 3,4$. Indeed, it can be shown \cite{IsPr} that $n$-split Casimir operators
\be\label{hcas}
\C_{(n)} = \sum_{i < j}^n \C_{ij} ,
\ee
where
\be\label{hcas1}
\C_{ij}=g^{ab} \left( I^{\otimes(i-1)}\otimes X_a \otimes
I^{\otimes(j-i-1)}\otimes X_b \otimes I^{\otimes(n-j)}\right) ,
\ee  
obey the characteristic identity
\be\label{main0}
\prod_{j} \left( \C_{(n)} +\lambda_j \right) =0 \; ,
\;\;\;\; (\lambda_i \neq \lambda_j \; \forall i \neq j) \; ,
\ee
where the product goes over 
the representations that appear in the expansion of $n$-power of the adjoint representation $\mathfrak{g}^{\otimes n}$ and we leave only the factors in which thr values $\lambda_j$  are pairwise different. Here 
\be 
\lambda_j= \frac{1}{2} (c_{2}^{(\Lambda)} -  n)
\ee
 is the (minus of the) eigenvalue of the $n$-split Casimir operator in the sub-representation $\Lambda$ in
$\mathfrak{g}^{\otimes n}$, $c_{2}^{(\Lambda)}$ is the value of the quadratic
Casimir operator in the representation $\Lambda$ (we use the normalization when $c_{2}^{(\mathfrak{g})}=1$).

Thus, if we know the corresponding powers of the $n$-split Casimir operators (or its invariant parts - symmetric, antisymmetric, "windows", hook, etc.), one can directly check  identity
\p{main0} and find all eigenvalues $\lambda_i$. Alas, the complexity of calculations increases rapidly for  higher  $n$-split Casimir operators. For example, for the symmetric part of $\ad^{\otimes 4}$, we have the characteristic identity of the 20th order \cite{AIKM}. Clearly,  it is rather an optimistic desire to go beyond $\ad^{\otimes 4}$ with this program.
The key problem is the necessity to calculate all the needed powers of the $n$-split Casimir operators before making any conclusion.

Fortunately enough, there is a possibility to drastically simplify all calculations. The main idea
can be formulated as follows:
\begin{itemize} 
	\item 
	The $n$-power of adjoint representations admits the decomposition
	\be
	\ad^{\otimes n} = \sum m_k R_k,
	\ee
where $R_k$ is the Casimir sub-representations in $\ad^{\otimes n}$ and $m_k$ denote their multiplicity
\item then the $n$-split Casimir operator obeys the equations
\be\label{eq1}
Tr \left( \C_{n}^k \right)  = \sum \left(\lambda_i\right)^k m_i \; dimR_i ,\quad k= 0,1,\ldots
\ee
Here $\lambda_i$ is the eigenvalue of the $n$-split Casimir operator on the representation $R_i$ and
$dimR_i$ denotes the dimension of  $R_i$ .
\item Clearly we have similar relations for the antisymmetric part of $\ad^{\otimes n}$, e.g.
\be \label{asymme}
\wedge \ad^{\otimes n} = \sum m_k {\tilde R_k} \quad \rightarrow \quad 
Tr \left(ASym_{n} \C_{n}^k \right)  = \sum \left(\lambda_i\right)^k m_i \; dim{\tilde R}_i .
\ee
\end{itemize}

The simplest example comes from  two basic equations that include the split 2-Casimir operator $ \C_{(2)}$ (see \p{tr2} )
\bea
Tr(ASym_2) & = & dimX_1+dimX_2, \label{eq21} \\
Tr(ASym_2 \C_{(2)}^k) & = & \left(-\frac{1}{2} \right)^k dimX_1 .\label{eq22}
\eea

In the next Section we will list equations \p{asymme} for the antisymmetric parts of 
$\ad^{\otimes 3}$ and $\ad^{\otimes 4}$. Finally we will find the  decomposition
of $\wedge\, \ad^{\otimes 5}$, which is our main result.

\setcounter{equation}0
\section{Decomposition of the antisymmetric parts of  $\ad^{\otimes 3}$ and $\ad^{\otimes 4}$}
To start with, we need to define the split 3,4-Casimir operators \cite{IsPr,IsKri} and their projections on the ant-symmetric parts:
\begin{itemize}
\item The split 3-Casimir operator reads :
\be
\C_{(3)}[i,j] =\left( C_{12} +C_{13} +C_{23} \right)\cE_3,
\ee
where 
\be
\cE_m =\prod_{k=1}^m \delta^{i_k}_{j_k}  \quad \mbox{and} \quad C_{mn} =C^{i_m i_n}_{j_m j_n}.
\ee
\item Correspondingly, the split 4-Casimir operator is:
\be
\C_{(4)}[i,j] = \left( C_{12}+C_{13} +C_{14} +C_{23} +C_{24}+C_{34} \right) \cE_4
\ee
\item The projectors on the antisymmetric parts have the following form:
\bea
ASym_3 & = & \frac{1}{3!}\left(1-\sigma_{23} +\sigma_{12}\sigma_{23}\right)\left( 1-\sigma_{12}\right)\cE_3, \label{asym3}\\
ASym_4 & = & \frac{1}{4!} \left(1-\sigma_{34}+\sigma_{23}\sigma_{34}-\sigma_{12}\sigma_{23}\sigma_{34}\right)\left(1-\sigma_{23} +\sigma_{12}\sigma_{23}\right)\left( 1-\sigma_{12}\right)\cE_4, \label{asym4} 
\eea
\end{itemize}
Additionally, we will also need some first traces of $Tr(ASym\C_{(3)})^k$ and $Tr(ASym\C_{(4)})^k$:
\begin{itemize}
\item {The split 3-Casimir operator:}
\bea
Tr(ASym_3) & = &\frac{1}{6} \dimg (\dimg-1) (\dimg-2) , \nn \\
Tr(ASym_3 \, \C_{(3)}) & = & -\frac{1}{2} \dimg  (\dimg-2), \nn \\
Tr(ASym_3 \, \C_{(3)}^2) & = & \frac{1}{4} \dimg^2 , \nn \\
Tr(ASym_3 \, \C_{(3)}^3) & = & -\frac{1}{8} \dimg  (\dimg+6)\label{tr3} .
\eea
\item{The split 4-Casimir operator:}
\bea
Tr(ASym_4) & = &\frac{1}{24} \dimg (\dimg-1) (\dimg-2) (\dimg-3) , \nn \\
Tr(ASym_4 \, \C_{(4)}) & = & -\frac{1}{4} \dimg  (\dimg-2) (\dimg-3), \nn \\
Tr(ASym_4 \, \C_{(4)}^2) & = & \frac{1}{8} \dimg (\dimg-3) (\dimg+4) , \nn \\
Tr(ASym_4 \, \C_{(4)}^3) & = & -\frac{1}{16} \dimg (\dimg-3) (\dimg+ 20)\label{tr4} .
\eea
\end{itemize}

\subsection{The basic equations}
\subsubsection{$\wedge \ad^{\otimes 3}$}
The antisymmetric part of $\ad^{\otimes 3}$ includes the following representations \cite{Vogel}:
\be\label{dec3}
\wedge \ad^{\otimes 3} = X_0\oplus X_2\oplus X_3\oplus Y_2\oplus Y_2'\oplus Y_2''.
\ee
The two basic equations that include the split 3-Casimir operator $\C_{(3)}$ read
\bea
Tr(ASym_3)  & = &  dimX_0+dimX_2+dimX_3+dimY_2+dimY_2'+dimY_2'', \label{eq31} \\
Tr(ASym_3 \C_{(3)}^k)  & = & \left(-\frac{3}{2}\right)^k dimX_0+\left(-\frac{1}{2} \right)^k dimX_2 +\left(-\frac{1}{2}-\halpha\right)^k dimY_2 +\left(-\frac{1}{2}-\hbeta\right)^k dimY_2'+ \nn \\
&& \left(-\frac{1}{2}-\hgamma\right)^k dimY_2'' ,\label{eq32}
\eea
where $\halpha,\hbeta$ and $\hgamma$ are the Vogel parameters normalized as 
$$
\halpha+\hbeta+\hgamma =\frac{1}{2} .
$$
The values of these parameters for simple Lie algebras are given in Vogel's Table 1.  

\begin{center}\label{tab1}
	Table 1. \\ [0.2cm]
	\begin{tabular}{|c|c|c|c|c|c|c|c|c|}
		\hline
		$\;\;$ & $sl(N)$ & $so(N)$&  $sp(N)$ &
		$\mathfrak{g}_2$ & $\mathfrak{f}_4$
		& $\mathfrak{e}_6$ &
		$\mathfrak{e}_7$ & $\mathfrak{e}_8$  \\
		\hline
		$\halpha$&\footnotesize $-1/N$ &\footnotesize $ -1/(N-2)$   &\footnotesize  $1/(N+2)$ &
		\footnotesize $-1/4$ &\footnotesize  $-1/9$ &\footnotesize  $-1/12$
		&\footnotesize  $-1/18$ &\footnotesize  $-1/30$  \\
		\hline
		$\hbeta$  &\footnotesize $1/N$ &
		\footnotesize $2/(N-2)$&\footnotesize $-2/(N+2)$ &
		\footnotesize $5/12$ &\footnotesize $5/18$ &\footnotesize $1/4$ &
		\footnotesize $2/9$ &\footnotesize $1/5$ \\
		\hline
		$\hgamma$  &\footnotesize 1/2& \footnotesize $(N-4)/(2N-4)$ &\footnotesize $(N+4)/(2N+4)$&
		\footnotesize $1/3$ &\footnotesize $1/3$ &\footnotesize $1/3$ &
		\footnotesize $1/3$ &\footnotesize $1/3$ \\
		\hline
	\end{tabular}
\end{center}
The eigenvalues of the Split 3-Casimir operator $ \C_{(3)}$ on the representations $X_0,X_2,X_3, Y_2,Y_2'$ and $Y_2''$ are
equal to $ -\frac{3}{2}, -\frac{1}{2},0,-\frac{1}{2}-\halpha,-\frac{1}{2}-\hbeta $ and $-\frac{1}{2}-\hgamma$, respectively.

\subsubsection{$\wedge \ad^{\otimes 4}$}
The antisymmetric part of $\ad^{\otimes 4}$ includes the following representations \cite{AIKM}
\be\label{dec4}
\wedge \ad^{\otimes 4} = X_1\oplus X_2\oplus X_3\oplus X_4 \oplus Y_2\oplus Y_2'\oplus Y_2'' \oplus B \oplus B' \oplus B'' \oplus C \oplus C' \oplus C''.
\ee
The two basic equations that include Split 4-Casimir operator $ \C_{(4)}$ read
\bea
Tr(ASym_4)  & = &  dimX_1+dimX_2+dimX_3+dimX_4+dimY_2+dimY_2'+dimY_2''+\nn \\
&& dimB+dimB'+dimB''+ dimC+dimC'+dimC'', \label{eq41} \\
Tr(ASym_4 \C_{(4)}^k)  & = & \left(-\frac{3}{2}\right)^k dimX_1+ \left(-1\right)^k dimX_2+ \left(-\frac{1}{2} \right)^k dimX_3 +  \nn \\
&& \left(-1-\halpha\right)^k dimY_2 +\left(-1-\hbeta\right)^k dimY_2'+ 
 \left(-1 -\hgamma\right)^k dimY_2''+ \nn \\
 &&\left(-1+\halpha\right)^k dimB +\left(-1+\hbeta\right)^k dimB'+  \left(-1+\hgamma\right)^k dimB''+ \nn \\
&& \left(-\frac{1}{2}-\frac{3}{2}\halpha\right)^k dimC +\left(-\frac{1}{2}-\frac{3}{2}\hbeta\right)^k dimC' + \left(-\frac{1}{2}-\frac{3}{2}\hgamma\right)^k dimC'' .\label{eq42}
\eea

\setcounter{equation}0
\section{Decomposition of the antisymmetric parts of  $\ad^{\otimes 5}$ }
One of the interesting peculiarities of the antisymmetric parts of $\ad^{\otimes n}$ was
noted in \cite{Mac}: it happened that among the representations entering into the decomposition of
$ASym\,\ad^{\otimes n}$ only one representation is new - $X_n$ and the rest of the representations already appeared in the decomposition of $ASym\,\ad^{\otimes (n-1)}$. In the paper \cite{Mac}, this
peculiarity was noted only for the exceptional algebras, but the structure of the representations in \p{dec2}, \p{dec3},  \p{dec4} confirms this property  for arbitrary Lie algebras, at least up to the decomposition of  $ASym\,\ad^{\otimes 4}$. In any case, one can try to check this idea at the level of  $ASym\,\ad^{\otimes 5}$ because all representations, their dimensions, and eigenvalues of the Casimir operator that appeared at the level of $ASym\,\ad^{\otimes 5}$ are known \cite{AIKM}.

The split 5-Casimir operator has the following structure:
\be
\C_{(5)}[i,j] = \left( C_{12} + C_{13} + C_{14} + C_{15}  +
C_{23}  +C_{24}  + C_{25}  +
C_{34}  + C_{35} +C_{45}\right) \cE_5 ,
\ee
while the projector on the antisymmetric part of $\C_{(9)}$ reads
\bea
ASym_5 &=& \frac{1}{5!} \left(1-\sigma_{45}+\sigma_{34}\sigma_{45} -\sigma_{23} \sigma_{34}\sigma_{45}+\sigma_{12}\sigma_{23}\sigma_{34}\sigma_{45}\right) \left(1-\sigma_{34} +\sigma_{23} \sigma_{34}-\sigma_{12}\sigma_{23}\sigma_{34}\right) \times \nn \\
&& \left(1-\sigma_{23}+\sigma_{12}\sigma_{23}\right)\left(1-\sigma_{12}\right) \cE_5. \label{asym5}
\eea
The first several  traces of $Tr(ASym_5 \, \C_{(5)}^k)$ can be easily calculates in a universal form:
\bea\label{meq5}
Tr(ASym_5) & = &\frac{1}{120} \dimg (\dimg-1)(\dimg-2)(\dimg-3)(\dimg-4) , \nn \\
Tr(ASym_5 \, \C_{(5)}) & = & -\frac{1}{12} \dimg (\dimg-2)(\dimg-3)(\dimg-4) , \nn \\
Tr(ASym_5 \, \C_{(5)}^2) & = & \frac{1}{24} \dimg (\dimg-3)(\dimg-4)(\dimg+10) , \nn \\
Tr(ASym_5 \, \C_{(5)}^3) & = & -\frac{1}{48} \dimg (\dimg-3)(\dimg^2+36 \dimg-124), \\
Tr(ASym_5 \, \C_{(5)}^4) & = & \frac{1}{96} \dimg (\dimg-3)(\dimg^2+114 \dimg-208+36(\dimg-10)(\dimg-9)\halpha \hbeta \hgamma). \nn
\eea

\subsection{$\wedge \ad^{\otimes 5}$ for the exceptional algebras}
The decomposition of the $\wedge \ad^{\otimes 5}$ has the simplest structure for the exceptional algebras because many of the representations have zero dimension for the exceptional algebras and thus they disappeared from the formulas.

The starting point is the general  equations \p{eq1}
\be
Tr \left(ASym_{5} \C_{(5)}^k \right)  = \sum \left(\lambda_i\right)^k m_i \; dim{\tilde R}_i .
\ee
These equations are naturally divided into two sets:
\begin{itemize}
	\item The first equation corresponds to $k=0$:
	\be\label{eq01} 
	Tr \left(ASym_{5}\right)  = \sum  m_i \; dim{\tilde R}_i
	\ee
	Here the set of representations ${\tilde R}_i$ includes all representations from Table 1 
	\p{table1} and the new representation $X_5$ with the eigenvalue $0$
	\item The set of equations with $k\geq 1$
	\be\label{eq02} 
	Tr \left(ASym_{5} \C_{(5)}^k \right)  = \sum \left(\lambda_i\right)^k m_i \; dim{\tilde R}_i .
	\ee
    Here the representations ${\tilde R}_i$ include only those appearing in the decomposition $\wedge \ad^{\otimes 4}$.
\end{itemize}
Now one can use the first several equations \p{eq02} to fix the multiplicities $m_i$
\bea
Tr(ASym_5 \C_{(5)}^k)   &=&  \left(-2\right)^k dimX_1+ 2 \left(-\frac{3}{2}\right)^k dimX_2+ \left(-\frac{1}{2} \right)^k dimX_4 + 2\left(-\frac{7}{6}\right)^k dimB'' + \nn \\
&& \left(-1-\frac{3}{2}\halpha \right)^k dimC +\left(-1-\frac{3}{2}\hbeta \right)^k dimC'+
\left(-\frac{1}{2}-\frac{3}{2}(\halpha+\hbeta)\right)^k dimE''  + \nn \\
&& \left(-\frac{1}{2}-2\halpha-\hbeta\right)^k dimF+  \left(-\frac{1}{2}-2\hbeta-\halpha\right)^k dimF' +  \nn \\
&& \left(-1 -3\halpha\right)^k dimY_3+ \left(-1-3\hbeta\right)^k dimY_3'+ \nn \\
&& \left(-\frac{1}{2}-2\halpha\right)^k dimI +\left(-\frac{1}{2}-2\hbeta\right)^k dimI' .\label{eq52}
\eea
Finally, the first equation \p{eq01}
\bea
Tr(ASym_5)  & = &  dimX_1+2 dimX_2+dimX_4+dimX_5+ 2 dimB''+\nn \\
&&  dimC+dimC'+dimE''+dimF+dimF'+ \nn \\
&& dimY_3+dimY_3'+dimI+dimI' ,\label{eq51} 
\eea
fixes the dimension of the representation $X_5$ to be
\be\label{dimx5}
dimX_5 = \frac{1}{5!} \dimg (\dimg-3)(\dimg-6)(\dimg^2-21 \dimg+8).
\ee
Thus, the decomposition of the antisymmetric part of $\ad^{\otimes 5}$ for the exceptional algebras reads
\be
\wedge\, \ad^{\otimes 5} = X_1\oplus 2 X_2\oplus X_4\oplus X_5 \oplus 2 B''  \oplus C \oplus C' 
 \oplus  E'' \oplus  F \oplus F'\oplus Y_3 \oplus Y_3' \oplus I \oplus I'.
\ee
The dimension of the representation $X_5$ \p{dimx5} was  suggested in \cite{Mac} for completely different arguments.

The component $X_5$ is rather interesting. For example, for the algebra $\mathfrak{g}_2$ we have a negative dimension of $X_5$ $dimX_5=-924$ which is compensated by the representation $I$ with eigenvalue $0$ and dimension $dimI|_{\mathfrak{g}_2} =924$. Note that $dimI'|_{\mathfrak{g}_2} =0$ and the last lines in \p{eq52} and \p{eq51} disappear together with the representation $X_5$. For  algebra $\mathfrak{f}_4$ the representation $X_5$ is reducible 
\be
dimX_5|_{\mathfrak{f}_4} =1 582308 = 629356 \oplus 952952,
\ee
while for the algebra $\mathfrak{e}_8$ the representation $X_5$ with the dimension
\be
dimX_5|_{\mathfrak{e}_8} =6899079264
\ee
is irreducible.

\subsection{Decomposition of the $\wedge\; \ad^{\otimes 5}$ for all algebras}
This case is slightly more complicated. The antisymmetric part of $\ad^{\otimes 5}$ includes the following representations:
\bea
\wedge \; \ad^{\otimes 5} &= &2 X_1\oplus 4 X_2 \oplus  \mathbb{M}_3 \oplus X_4\oplus X_5 \oplus 3 B  \oplus 3 B' \oplus 3 B'' \oplus 2 C \oplus 2 C' \oplus 2 C'' \oplus \nn \\
&&  E \oplus  E' \oplus  E'' \oplus  F \oplus F'\oplus F''\oplus F'''\oplus F^{(4)}\oplus F^{(5)} \oplus \nn \\
&&  Y_3 \oplus Y_3'  \oplus Y_3''\oplus I \oplus I'\oplus I'',
\eea
where
\be\label{M3}
dim\mathbb{M}_3 = \left\{ \begin{array}{l} 
	\frac{1}{6} \left(N^2-1\right) \left(N^2-2\right) \left(N^2-9\right) \quad \mbox{for }SL(N) ,\\
    \frac{1}{72} N(N^2-1)(N-3)(N^2-16) \quad \mbox{for } SO(N) ,\\
	0 \quad \mbox{for exceptional algebras}.
	\end{array} \right.
\ee
The universal dimension of  $\mathbb{M}_3$ reads\footnote{The universal dimension 
	$dim\mathbb{K}_3$ discussed in \cite{AIKM}.}
\be
dim\mathbb{M}_3 =dim\mathbb{K}_3-dimX_3-\frac{1}{2}(\dimg+3)\frac{dimB\, dimB'\, dimB''}{dimY_2\, dimY_2'\, dimY_2''}.
\ee

The two basic equations that include the split 5-Casimir operator $ \C_{(5)}$ read
\bea
Tr(ASym_5)  & = &  2\, dimX_1+4\, dimX_2+dim\mathbb{M}_3+dimX_4+dimX_5+ 3\, dimB+3\, dimB'+3\, dimB''+\nn \\
&& 2 dimC+ 2 dimC'+ 2dimC''+ dimE+dimE'+dimE''+ \nn \\
&& dimF+dimF'+dimF''+dimF'''+dimF^{(4)}+dimF^{(5)}+ \nn \\
&& dimY_3+dimY_3'+dimY_3''+dimI+dimI'+dimI'' ,\label{eq51a} \\
Tr(ASym_5 \C_{(5)}^k)  & = & 2 \left(-2\right)^k dimX_1+ 4 \left(-\frac{3}{2}\right)^k dimX_2+ \left( -1\right)^k dim\mathbb{M}_3+\left(-\frac{1}{2} \right)^k dimX_4 + \nn \\
&& 3\left(-\frac{3}{2}+\halpha \right)^k dimB +3\left(-\frac{3}{2}+\hbeta \right)^k dimB' +3\left(-\frac{3}{2}+\hgamma \right)^k dimB'' + \nn \\
&& 2\left(-1-\frac{3}{2}\halpha \right)^k dimC +2\left(-1-\frac{3}{2}\hbeta \right)^k dimC' +2\left(-1-\frac{3}{2}\hgamma \right)^k dimC'' + \nn \\
&& \left(-\frac{1}{2}-\frac{3}{2}(\hbeta+\hgamma)\right)^k dimE +\left(-\frac{1}{2}-\frac{3}{2}(\halpha+\hgamma)\right)^k dimE' +\left(-\frac{1}{2}-\frac{3}{2}(\halpha+\hbeta)\right)^k dimE''  + \nn \\
&& \left(-\frac{1}{2}-2\halpha-\hbeta\right)^k dimF+  \left(-\frac{1}{2}-2\hbeta-\halpha\right)^k dimF' + \left(-\frac{1}{2}-2\hgamma-\hbeta\right)^k dimF''+  \nn \\
&& \left(-\frac{1}{2}-2\halpha-\hgamma\right)^k dimF'''+ \left(-\frac{1}{2}-2\hbeta-\hgamma\right)^k dimF^{(4)}+ \left(-\frac{1}{2}-2\hgamma-\halpha\right)^k dimF^{(5)}+ \nn \\
&& \left(-1 -3\halpha\right)^k dimY_3+ \left(-1-3\hbeta\right)^k dimY_3'+ \left(-1-3\hgamma\right)^k dimY_3''+ \nn \\
&& \left(-\frac{1}{2}-2\halpha\right)^k dimI +\left(-\frac{1}{2}-2\hbeta\right)^k dimI'
+\left(-\frac{1}{2}-2\hgamma\right)^k dimI'' .\label{eq52a}
\eea

\setcounter{equation}0
\section{Conclusion}
In the present paper, using the  split Casimir operators we  have
found the decomposition of the antisymmetric part of $\ad^{\otimes 5}$. As it was suggested in \cite{Mac}, the decomposition contains the representations that appeared in the decomposition of 
 $\ad^{\otimes 4}$ and only one new representation $X_5$. The dimension of this representation is given by the same expression \p{dimx5}, as it was proposed in \cite{Mac}.  Our decomposition is valid for all Lie algebras.
 
 The new feature of the decomposition we found consists in the multiplicities different from 1 and 0 for some representations. Another peculiarity is the structure of the representation
 $\mathbb{M}_3$ \p{M3} (modified $X_3$).
 
 We would like to stress that the use of the main formulas \p{eq1} is preferable as compared to the characteristic identity. All we need to know to find decomposition formulas is
 the traces of powers of the split Casimir operators and their eigenvalues on the given representations. Alas, starting from the fourth power, traces of the split Casimir operators
 cannot be expressed only in terms of $\dimg$. They still have a universal form, but 
 new terms of the form $\left( \halpha \hbeta\hgamma\right)^k  Poly\left[\dimg\right]$ appear systematically. If a generic expression, for example of  $Tr(ASym_n \C_{(n)}^k)$ were found,  then the general proof of  decomposition formulas can be derived.
 
 \section*{Acknowledgments}
 The work of API was supported by the RNF, grant  23-11-00311.
 
 SOK acknowledges partial financial support of the Ministry of Science and Higher Education of Russia, Government Order  for 2023-2025, Project No. FEWM-2023-0015 (TUSUR).

\def\theequation{A.\arabic{equation}}
\setcounter{equation}0
\section*{Appendix}
In the antisymmetric part of $\ad^{\otimes 4}$, for the exceptional algebras the following representations appeared

\begin{center}
{\bf	Table 2} 
\end{center}
\be\label{table1}
\begin{tabular}{|c|c|c|c|c|} 
	\hline
Reps & $\C^{(4)}$ & $\C^{(2)} $ &  $\C^{(5)} $ & dim  \\ \hline
$X_0$ & -2 & 0 & $ - \frac{5}{2}$ & 1 \\
$X_1$ &$-\frac{3}{2}$ & 1 &  -2 & $\dimg $\\
$X_2$ & -1 & 2& $-\frac{3}{2}$ & $ \frac{1}{2} \dimg (\dimg-3)$ \\
$X_3$ & $-\frac{1}{2} $ & 3 & -1 & $\frac{1}{6} \dimg(\dimg-1)(\dimg-8)$ \\
$X_4$ & 0 & 4 & $-\frac{1}{2}$ & $\frac{1}{24} \dimg(\dimg-1)(\dimg-3)(\dimg-14)$ \\
$Y_2$ & $-1-\halpha$ &$-2\halpha+2 $ & $-\frac{3}{2} -\halpha$&$ -\frac{5(3\halpha-2)(6\halpha+5)}{\halpha^2(6\halpha-1)(12\halpha-1)}$ \\
$Y_2'$ & $-1-\hbeta$ & $-2\hbeta+2$& $-\frac{3}{2}-\hbeta$& 
$-\frac{270(\halpha-1)(2\halpha+1)}{\halpha (6\halpha-1)^2(12\halpha-1)} $\\
$B''$ & $-1+\hgamma$ & $2\hgamma+2 $ & $-\frac{3}{2}+\hgamma$ &
$-\frac{27 (\halpha -1) (2 \halpha +1) (3 \halpha -2) (3 \halpha +2) (6 \halpha -5) (6 \halpha +5)}{\halpha ^2 (6 \halpha -1)^2 (9 \halpha -1)
	(18 \halpha -1)}$ \\
$C$ & $ -\frac{1}{2} -\frac{3}{2} \halpha$ & $-3\halpha+3$ & $-1 -\frac{3}{2}\halpha$ &
$\frac{40 (2 \halpha -1) (2 \halpha +1) (6 \halpha -5) (6 \halpha +5)}{3 \halpha ^3 (6 \halpha -1) (9 \halpha -1) (12 \halpha -1)}$ \\
$C'$ & $-\frac{1}{2}-\frac{3}{2}\hbeta$ & $-3\hbeta+3$ & $-1 -\frac{3}{2}\hbeta$ &
$\frac{5120 (\halpha -1) (3 \halpha -2) (3 \halpha +1) (3 \halpha +2)}{\halpha  (6 \halpha -1)^3 (12 \halpha -1) (18 \halpha -1)}$ \\
$J''$ & $-1+2\hgamma$ & $4\hgamma+2$ & $-\frac{3}{2}+2\hgamma$ &
$\frac{729 (\halpha -1) (2 \halpha -1) (2 \halpha +1) (3 \halpha -2) (3 \halpha +1) (4 \halpha +1) (6 \halpha +5) (12 \halpha -5)}{\halpha ^2 (6
	\halpha -1)^2 (9 \halpha -1) (12 \halpha -1)^2 (18 \halpha -1)}$ \\
$Y_4$ & $-6\halpha$ & $-12\halpha+4 $& $-\frac{1}{2} -6\halpha$ &
$-\frac{5 (\halpha -1) (2 \halpha -1) (3 \halpha -2) (6 \halpha -5) (6 \halpha +5) (7 \halpha -1) (9 \halpha -2) (12 \halpha -5)}{6 \halpha ^4
	(6 \halpha -1)^2 (9 \halpha -1) (12 \halpha -1) (18 \halpha -1) (24 \halpha -1)}$ \\
$Y_4'$ & $-6\hbeta$ & $ -12 \hbeta+4$ & $-\frac{1}{2} -6\hbeta$ &
$-\frac{5 (\halpha -1) (2 \halpha +1) (3 \halpha +1) (3 \halpha +2) (4 \halpha +1) (6 \halpha +5) (18 \halpha +1) (42 \halpha -1)}{\halpha ^2 (6
	\halpha -1)^4 (8 \halpha -1) (9 \halpha -1) (12 \halpha -1) (18 \halpha -1)}$\\
$D$ & $-3\halpha-\hbeta$ &$-6\halpha-2\hbeta+4$ & $-\frac{1}{2} -3\halpha-\hbeta$ &
$-\frac{270 (\halpha -1) (2 \halpha -1) (2 \halpha +1) (3 \halpha +2) (5 \halpha -1) (6 \halpha +5) (12 \halpha -5)}{\halpha ^3 (6 \halpha -1)^3
	(12 \halpha -1)^2 (24 \halpha -1)}$ \\
$D'$ & $-3\hbeta-\halpha $ & $-6\hbeta-2\halpha +4$ & $-\frac{1}{2} -3\hbeta-\halpha$&
$-\frac{10 (\halpha -1) (3 \halpha -2) (3 \halpha +1) (4 \halpha +1) (6 \halpha -5) (6 \halpha +5) (30 \halpha +1)}{\halpha ^3 (6 \halpha -1)^3
	(8 \halpha -1) (12 \halpha -1)^2}$ \\
$E''$ & $-\frac{3}{4}+\frac{3}{2}\hgamma$ & $3\hgamma+\frac{5}{2} $ & $-\frac{5}{4}+\frac{3}{2}\hgamma$ &
$\frac{2048 (\halpha -1) (2 \halpha -1) (2 \halpha +1) (3 \halpha -2) (3 \halpha +1) (3 \halpha +2) (6 \halpha -5) (6 \halpha +5)}{\halpha ^2 (6
	\halpha -1)^2 (8 \halpha -1) (12 \halpha -1)^2 (24 \halpha -1)}$ \\
$H$ & $-3\halpha$ & $-6\halpha+4$& $-\frac{1}{2}-3\halpha $ & 
$-\frac{5 (\halpha -1) (2 \halpha +1) (3 \halpha +2) (5 \halpha -1) (6 \halpha -5) (6 \halpha +5) (9 \halpha -2) (12 \halpha -5)}{3 \halpha ^4
	(6 \halpha -1)^3 (9 \halpha -1) (12 \halpha -1)^2}$ \\
$H'$ & $-3\hbeta$ &$-6\hbeta+4$ & $-\frac{1}{2}-3\hbeta $ &
$-\frac{10 (\halpha -1) (3 \halpha -2) (3 \halpha +2) (4 \halpha +1) (6 \halpha -5) (6 \halpha +5) (18 \halpha +1) (30 \halpha +1)}{3 \halpha ^3
	(6 \halpha -1)^4 (12 \halpha -1)^2 (18 \halpha -1)}$ \\
$G$ & $- 4 \halpha$ & $ -8\halpha+4 $ & $-\frac{1}{2}-4\halpha$ &
$\frac{5 (\halpha -1) (2 \halpha +1) (3 \halpha -2) (6 \halpha -5) (6 \halpha +5) (9 \halpha -2) (12 \halpha -5)}{2 \halpha ^4 (6 \halpha -1)^2
	(12 \halpha -1)^2 (18 \halpha -1)}$ \\
$G'$ & $-4\hbeta$ & $-8\hbeta+4 $& $-\frac{1}{2}-4\hbeta$& 
$\frac{135 (\halpha -1) (2 \halpha +1) (3 \halpha -2) (3 \halpha +2) (4 \halpha +1) (6 \halpha +5) (18 \halpha +1)}{\halpha ^2 (6 \halpha -1)^4
	(9 \halpha -1) (12 \halpha -1)^2}$ \\
$F$ & $-2\halpha-\hbeta$ & $-4\halpha-2\hbeta+4$ & $-\frac{1}{2}-2\halpha-\hbeta$& 
$-\frac{270 (\halpha -1) (2 \halpha +1) (3 \halpha -2) (3 \halpha +1) (6 \halpha +5) (12 \halpha -5)}{\halpha ^3 (6 \halpha -1)^2 (12 \halpha
	-1)^2 (18 \halpha -1)}$ \\
$F'$ & $ -2\hbeta-\halpha$ & $ -4\hbeta-2\halpha+4$ &$-\frac{1}{2}-2\hbeta-\halpha$ &
$-\frac{3645 (\halpha -1) (2 \halpha -1) (2 \halpha +1) (3 \halpha -2) (4 \halpha +1) (6 \halpha +5)}{\halpha ^2 (6 \halpha -1)^3 (9 \halpha -1)
	(12 \halpha -1)^2}$ \\
$Y_3$ & $-\frac{1}{2} - 3\halpha $ & $-6\halpha+3$  &$-1-3\halpha$& 
$-\frac{10 (\halpha -1) (3 \halpha -2) (5 \halpha -1) (6 \halpha -5) (6 \halpha +5)}{3 \halpha ^3 (6 \halpha -1)^2 (12 \halpha -1) (18 \halpha
	-1)}$ \\
$Y_3'$ &$-\frac{1}{2} - 3\hbeta $ & $-6\hbeta+3$ &$-1-3\hbeta$ & 
$-\frac{5 (\halpha -1) (2 \halpha +1) (3 \halpha +2) (6 \halpha +5) (30 \halpha +1)}{\halpha ^2 (6 \halpha -1)^3 (9 \halpha -1) (12 \halpha -1)}$ \\
$I$ & $-2\halpha$ & $-4\halpha+4$ & $-\frac{1}{2} - 2\halpha $& 
$-\frac{5 (3 \halpha -2) (3 \halpha +1) (3 \halpha +2) (5 \halpha -1) (6 \halpha -5) (6 \halpha +5) (12 \halpha -5)}{2 \halpha ^4 (6 \halpha
	-1)^2 (8 \halpha -1) (9 \halpha -1) (12 \halpha -1)}$\\
$I'$ & $-2\hbeta$ & $-4\hbeta+4$ & $-\frac{1}{2} - 2\hbeta$&
$-\frac{3645 (\halpha -1) (2 \halpha -1) (2 \halpha +1) (3 \halpha +2) (4 \halpha +1) (6 \halpha -5) (30 \halpha +1)}{\halpha ^2 (6 \halpha
	-1)^4 (12 \halpha -1) (18 \halpha -1) (24 \halpha -1)}$  \\ 
\hline
\end{tabular}
\ee


\begin{thebibliography}{99}
\addtolength{\itemsep}{-1pt}

\bibitem{Mac} A.J.~Macfarlane, H.~Pfeiffer, {\textit Representations of the exceptional and other Lie algebras with integral eigenvalues of the Casimir operator}, 
J.~Phys.~A:~ Math.~ Gen. {\bf 36} (2003) 2305;
{\tt arXiv:math-ph/0208014}

\bibitem{Vogel} P.~Vogel, \textit{The universal Lie algebra}, preprint 1999,
https://webusers.imj-prg.fr/\~{}pierre.vogel/grenoble-99b.pdf .

\bibitem{Cohen} A.M.~Cohen, R.~de~Man, \textit{Computational evidence for Deligne's conjecture regarding exceptional Lie groups}, C.R.~Acad.~Sci.~Paris {\bf 322} (1996) 427.

\bibitem{Landsberg} J.M.~Landsberg, L.~Manivel, \textit{A universal dimension formula for complex simple Lie algebras}, Adv. Math. {\ bf 201} (2006) 379.

\bibitem{Deligne} P.~Deligne, \textit{ La s\'{e}rie exceptionnelle des groupes de Lie}, C.R.~Acad.~Sci. {\bf 322} (1996) 321. 

\bibitem{IsPr} A.P.~Isaev, and A.A.~Provorov,
 \textit{Projectors on invariant subspaces of representations
   $\ad^{\otimes 2}$ of Lie algebras $so(N)$ and $sp(2r)$ and Vogel parametrization}, TMF {\bf 206} (2021) 3; {\tt arXiv:2012.00746[math-ph]}.

\bibitem{AIKM} M.~Avetisyan, A.P.~Isaev, S.O.~Krivonos, R.~Mkrtchyan,
{\textit The uniform structure of $g^{\otimes 4}$}, 
{\tt arXiv:2311.05358[math-ph]}.

\bibitem{IsKri} A.P.~Isaev, and S.O.~Krivonos,
 \textit{Split Casimir operator for simple Lie algebras, solutions of Yang-Baxter equations and Vogel parameters},  Journal of Mathematical Physics {\bf 62} (2021) 083503; 
{\tt arXiv:2102.08258[math-ph]}.
 
 \bibitem{IsKri2} A.P.~Isaev, S.O.~Krivonos, A.A.~Provorov,
{\textit Split Casimir operator for simple Lie algebras in the cube of ad-representation and Vogel parameters},
 Int.~J.~Mod.~Phys. A {\bf 38} (2023) 235003; {\tt arXiv:2212.14761[math-ph]}.

 \bibitem{Book2} A.P.~Isaev, V.A.~Rubakov,
\textit{Theory Of Groups And Symmetries: Finite Groups, Lie Groups, And Lie Algebras}, World Scientific (2018).

\end{thebibliography}
\end{document}